\documentclass[prl,twocolumn,floatfix,superscriptaddress,aps]{revtex4-2}

\usepackage[utf8]{inputenc}
\usepackage[T1]{fontenc}

\usepackage{graphicx}
\usepackage{bm}

\usepackage{amsmath}
\usepackage{units}

\usepackage{hyperref}
\hypersetup{colorlinks=true, citecolor=blue, urlcolor=blue, linkcolor=blue}

\renewcommand{\vec}[1]{\boldsymbol{#1}}

\begin{document}
\title{Enhanced magnetic dichroism in darkfield UV photoemission electron microscopy}
\author{Maximilian Paleschke}
\author{David Huber}
\author{Friederike W\"{u}hrl}
\affiliation{Institute of Physics, Martin Luther University Halle-Wittenberg, D-06099 Halle (Saale), Germany}

\author{Cheng-Tien Chiang}
\affiliation{Institute of Atomic and Molecular Sciences, Academia Sinica, Taipei, Taiwan}

\author{Frank O. Schumann}
\affiliation{Max-Planck-Institut f\"{u}r Mikrostrukturphysik, 06120 Halle, Germany}

\author{J\"{u}rgen Henk}
\author{Wolf Widdra}
\affiliation{Institute of Physics, Martin Luther University Halle-Wittenberg, D-06099 Halle (Saale), Germany}

\email{wolf.widdra@physik.uni-halle.de}

\date{\today}

\begin{abstract}
    Photoemission electron microscopy (PEEM) has evolved into an indispensable tool for structural and magnetic characterization of surfaces at the nanometer scale. In strong contrast to synchrotron-radiation-based X-ray PEEM as a leading method for element-specific magnetic properties via magnetic circular dichroism (MCD), laboratory ultraviolet (UV) PEEM has seen limited application with much smaller dichroic effects for in-plane magnetization. Here we introduce darkfield PEEM as a novel approach to enhance MCD contrast in threshold photoemission, enabling efficient MCD imaging with significantly enhanced contrast by an order-of-magnitude for Fe(001). This advancement paves the way for
    MCD imaging on femtosecond timescales using modern lasers. The experimental results will be quantitatively benchmarked against advanced relativistic photoemission calculations.
\end{abstract}


\maketitle
Ultrafast spin and magnetization dynamics are rapidly growing fields in condensed matter physics, holding great promise for both fundamental research and future device applications. Ultrafast imaging of magnetic domains on the micrometer scale is well established using all-optical methods, such as magneto-optical Kerr microscopy, which is inherently limited by optical diffraction. 
Alternatively, ultrafast scanning tunneling microscopy (STM) offers spatial resolution down to the atomic scale. However, its sensitivity to electron spin dynamics at surfaces depends on the spin-polarized current from the tip \cite{yoshida14}, and the recent development of ultrafast STM using THz electric fields \cite{mueller24} only provides the spin information if spin-orbit split electronic states can be probed in the tunneling process \cite{roelcke24}. In contrast, the combination of ultrafast lasers with the Lorentz transmission electron microscopy can offer the direct magnetic contrast in the femtosecond time domain in the transmitted volume \cite{daSilva18, moeller20, tengdin20}.

To access the surface-sensitive magnetization dynamics, magnetic circular dichroism (MCD) provides a known contrast mechanism used for imaging magnetic domains in photoelectron emission microscopy (PEEM). The intensity recorded for a particular domain changes with the helicity of the incident radiation, thereby producing magnetic contrast without need for explicitly detecting the electron spin. By tuning the incident X-ray radiation to a magnetic core level absorption edge, substantial and element-specific MCD asymmetries have been reported. With the wide availability of tunable synchrotron radiation, this technique of XMCD-PEEM is well established for magnetic domain imaging on the nanometer scale \cite{kuch15}. However, the pulse length of synchrotron radiation of typically $\unit[30-50]{ps}$ renders XMCD-PEEM unsuitable on ultrafast timescales. 
The reduced pulse length of X-ray free electron lasers (XFEL) can solve this issue \cite{Kutnyakhov20}. However, the general availability is lower than for lab-based experiments. Lower pulse repetition rates and time-restricted beamtimes of the XFEL limit available photoemission statistics. Here, we have combined PEEM with the MHz repetition rates of a high-power laser system as well as a Hg discharge lamp, thereby minimizing the space-charge effects and allowing extended experiments in the laboratory. UV laser sources excite electrons from close to the Fermi level to energies slightly above the escape threshold. The reported MCD contrasts, especially for in-plane magnetization, are so small in threshold photoemission \cite{marx2000} that UV-PEEM has been discarded for magnetic domain imaging in the last two decades. Obviously, magnetic contrast needs to be increased for domain imaging with ultrashort laser pulses.

As we demonstrate here, the concept of darkfield PEEM in threshold photoemission allows efficient MCD imaging with an order-of-magnitude enhanced MCD contrast for in-plane magnetization. It paves the way for MCD imaging on femtosecond timescales with modern UV laser sources. Darkfield PEEM imaging uses an aperture for photoelectron momentum selection in the back focal plane of the electron imaging column prior to forming the real-space image. We will demonstrate this for the in-plane magnetic structure at the Fe(001) surface and compare quantitatively the experimental results with fully relativistic photoemission calculations.

Following initial reports of magnetic dichroism in UV photoemission and its theoretical description in the 1990s \cite{schneider1991,henk1996,feder1996}, Marx \textit{et al.}\ reported the first observation of magnetic dichroism in threshold PEEM in 2000 \cite{marx2000}. This study of polycrystalline Fe revealed an asymmetry in magnetic linear dichroism of $\unit[0.37]{\%}$. Subsequent spectroscopic studies confirmed the presence of both circular and linear dichroism in various ferromagnetic materials. Building on Marx's experimental work, Nakagawa \textit{et al.}\ studied Ni films adsorbed with Cs and discovered significant asymmetries of up to $\unit[12]{\%}$ in circular dichroism PEEM for out-of-plane magnetized domains \cite{nakagawa2006,nakagawa2007,nakagawa2009,nakagawa2012}. This work also demonstrated the feasibility of using pulsed laser light for dichroism imaging. However, due to the limited photon energy range of common optical laser setups, Cs remained necessary in most photoemission experiments in order to reduce the work function \cite{kronseder2011, meier2017}, although PEEM studies using a deep-UV laser with a photon energy of $\unit[7]{eV}$ have been reported \cite{zhao2019}. 

The theoretical framework for valence-band dichroism was primarily developed in the 1990s and early 2000s \cite{tamura1987, feder1996,henk1996,kuch1996a,feder1998,venus1994,venus1997,kuch2001} and was bolstered by pioneering experiments \cite{venus95, hild2008, hild2009, hild2010}. It is based on calculating the relativistic electronic  structures in conjunction with a theoretical description of the photoemission process. 
The extension of this description to threshold photoemission predicted that experimentally accessible magnetic dichroism levels are expected \cite{feder1998}.

\begin{figure}
    \centering
    \includegraphics[width = \columnwidth]{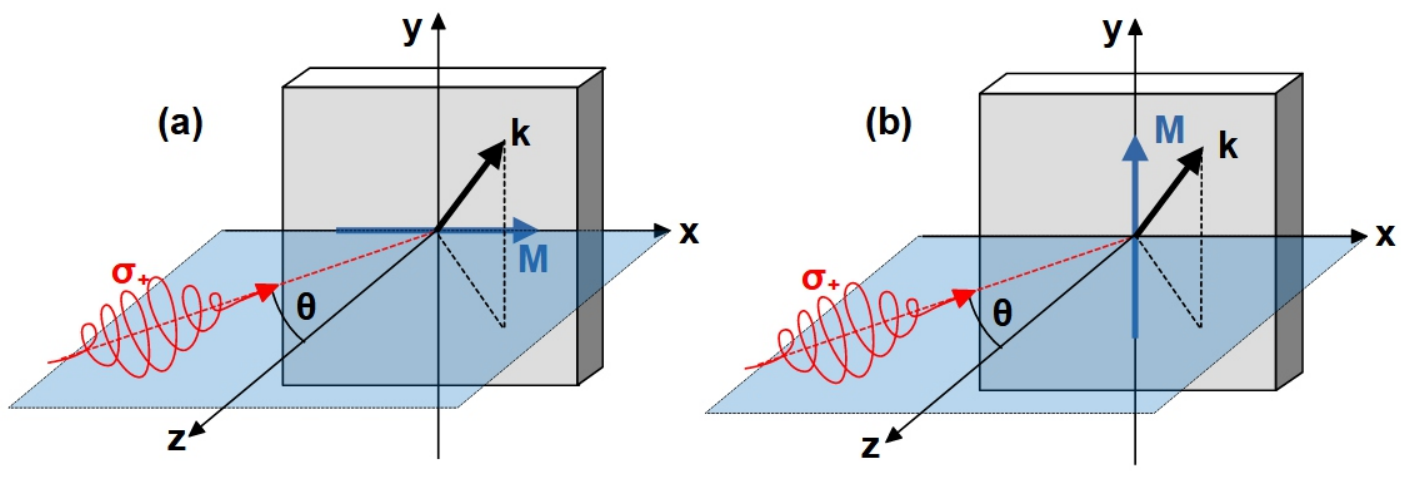}
    \caption{Symmetry analysis. A circular polarized laser pulse (shown in red, with helicity $\sigma_{+}$ impinges onto a magnetic domain (rectangular solid). The light incidence direction and the surface normal ($z$-axis) span the scattering plane (blue; $xz$-plane) with the magnetization direction $\vec{M}$ oriented within (a) or perpendicular (b) to the scattering plane, respectively. The off-normal detection of photoelectrons with wavevector $\vec{k}$ (black arrow) results in a chiral setup.}
    \label{fig:symmetry}
\end{figure}

For a simplified conceptual approach, we consider a surface with fourfold symmetry, as e.g. the (001) fcc or bcc surfaces with magnetic easy axes along one of the four [100] 
or [110] 
directions. Let us assume light incidence along the surface normal (the case of $\theta = 0$ in Fig.~\ref{fig:symmetry}). The photoemission intensity of electrons detected with off-normal wavevector $\vec{k}$ depends then on the helicity, $\sigma_{+}$ or $\sigma_{-}$, of the incident circularly polarized laser radiation and on the two orientations $\pm M$ of the in-plane magnetization in a selected domain, yielding four intensities $I_{\vec{k}}(\sigma_{\pm}, \pm M)$ (shortened $I_{\pm \pm}$). The latter intensities are combined into the total intensity
\begin{align}
    I & \equiv I_{+ +} + I_{+ -} + I_{- +} + I_{- -}. 
\end{align}
In order to disentangle the two main contrast mechanisms, we define appropriate asymmetries \cite{henk1998},
\begin{subequations}
\begin{align}
    A_{\mathrm{pol}} & \equiv \left[ \left( I_{+ +} + I_{+ -} \right) - \left( I_{- +} + I_{- -} \right) \right] / I,
    \label{eq:Apol}
    \\
    A_{\mathrm{ex}} & \equiv \left[ \left( I_{+ +} + I_{- -} \right) - \left( I_{+ -} + I_{- +} \right) \right] / I.
    \label{eq:Aex}
\end{align}    
\end{subequations}
In the polarization asymmetry $A_{\mathrm{pol}}$ the magnetization's orientation is averaged out; it thus encodes contrast due to the light's helicity, as if the domain were nonmagnetic. Contrast due to the exchange splitting is quantified by the exchange asymmetry $A_{\mathrm{ex}}$, in which one averages over the mutual orientations of helicity and magnetization. Note that the \emph{chiral geometry} for photoelectrons with \emph{off-normal} wavevector $\vec{k}$ outside the scattering plane results in magnetic dichroism and, hence, in magnetic contrast.

If the scattering plane is a mirror plane of the lattice, the photoemission intensities for fixed $\vec{k}$ within the scattering plane obey $I_{+ +} = I_{- -}$ for a magnetization within the scattering plane (Fig.~\ref{fig:symmetry}(a)). This results in a nonzero $A_{\mathrm{ex}}$ for $\vec{k}\neq 0$, but vanishing $A_{\mathrm{pol}}$. For magnetization perpendicular to the scattering plane, $I_{+ +} = I_{- +}$ holds that leads to vanishing $A_{\mathrm{pol}}$ and vanishing $A_{\mathrm{ex}}$.

In the following, we compare experimental results for a photon energy of $\unit[5.2]{eV}$ with theoretical MCD asymmetries based on relativistic photoemission computations for Fe(001) using the computer program package \textsc{omni}. 
The latter is based on the spin-polarized relativistic layer Korringa-Kohn-Rostoker method as applied earlier e.g. in Ref.~\onlinecite{tusche2018} and references therein. The sample is taken as semi-infinite Fe(100).

Self-consistent electronic-structure calculations have been performed using the local spin-density approximation for the exchange-correlation functional in density-functional theory. For solving the single-site scattering problem, the layer-KKR method uses an expansion of the scattering solutions with respect to angular momentum, in this work up to $l_{\mathrm{max}} = 3$. The potentials of the seven outermost layers of the semi-infinite system differ from that of the remainder (bulk). The interlayer scattering relies on a plane-wave expansion, here with at least 45~plane waves. For the image-potential barrier we take a smooth shape described in Ref.~\onlinecite{henk1993}. The calculated bulk and the surface electronic structures as well as the layer-resolved magnetic moments agree with those computed and published elsewhere. 

The results of the electronic-structure calculations serve as basis for the spin- and angle-resolved photoemission calculations, for which the same potentials as for the electronic-structure calculations are used. The spin-polarized photocurrent is calculated within the one-step model of photoemission, taking a time-reversed LEED state as final state. Photoelectrons excited within the topmost $50$~layers are considered in order to obtain converged intensities and spin polarizations. The calculated spin-density matrix of the photoelectron allows to derive all components of the photoelectron's spin polarization vector.

The experimental photocurrent has been recorded for $\unit[65]{^{\circ}}$ grazing light incidence within the [100] high-symmetry direction in a standard PEEM setup (Focus GmbH, H\"{u}nstetten). As light source either a mercury discharge lamp or the frequency-doubled output of a non-collinear optical amplifier (NOPA) with circular polarization optics is used \cite{duncker2012,gillmeister2020, paleschke2021}. The Fe(001) surface has been prepared by standard surface science procedures (as sputtering and annealing) and confirmed by low-energy electron diffraction and Auger or X-ray photoelectron spectroscopy.

\begin{figure}
    \centering
    \includegraphics[width = 0.7\columnwidth]{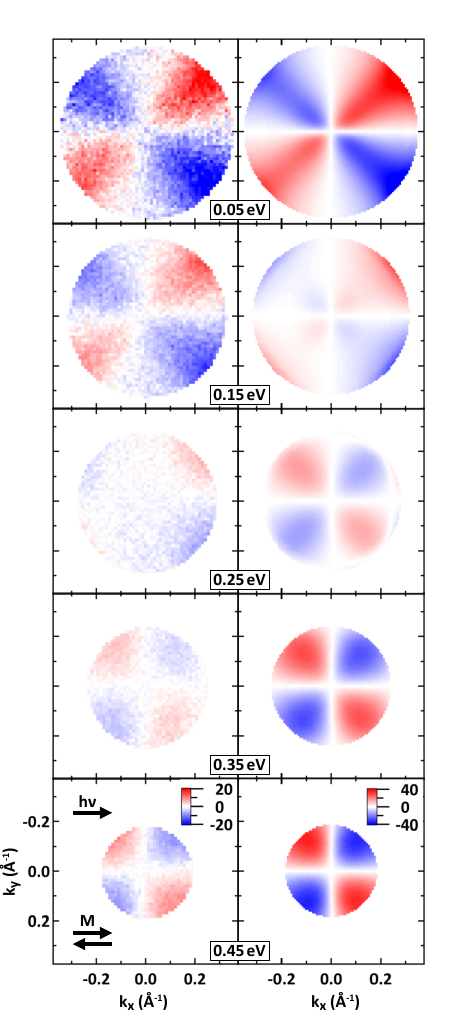}
    \caption{Momentum-resolved polarization asymmetry $A_{\mathrm{pol}}$ of Fe(001) at selected binding energies for $\unit[65]{^{\circ}}$ grazing light incidence. Left column: experimental results. The arrow marked $h \nu$ indicates the light incidence direction. The arrows $\textsf{M}$ represent the two magnetization directions considered for $A_{\mathrm{pol}}$. Right column: respective theoretical results obtained from photoemission calculations. The binding energy is indicated at each panel. The color scale, showing $A_{\mathrm{pol}}$ as defined in Eq.~\eqref{eq:Apol} in percent, is identical for all panels in a column. 
    }
    \label{fig:Apol}
\end{figure}

The $\vec{k}_{\parallel}$-dependent pattern of the polarization asymmetry $A_{\mathrm{pol}}$, defined in Eq.~\eqref{eq:Apol} and depicted in Fig.~\ref{fig:Apol}, depends on the binding energy of the initial states. Both experimental (left column) and theoretical data (right column) show that this contrast mechanism is sizable with absolute values up to about $\unit[20]{\%}$ in experiment and $\unit[40]{\%}$ in theory; it can thus hardly be ignored. 

The theoretical pattern in the momentum space (right column in Fig.~\ref{fig:Apol}) exhibits a nodal line at $k_{y} = 0$ and a nodal line at almost $k_{x} = 0$. Moreover, one finds a change of sign if $k_{y}$ is reversed. These features are imposed by the symmetry of the setup. Note that an anti-symmetric pattern with respect to the $k_{x} = 0$ \textit{and} $k_{y} = 0$ lines follows strictly only for normal light incidence \cite{schumann2024}. However, the breaking of the anti-symmetric behavior with respect to the $k_{x} = 0$ line due to the off-normal light incidence is hardly visible. The experimental data (left column) display the same features, and the overall agreement between experiment and theory is remarkably good, which includes also the sign change for binding energies above and below 0.2\,eV. Note that the experimental asymmetries have been determined from two independent sets of 2D momentum maps for magnetization directions $+\vec{M}$ and $-\vec{M}$ oriented along the +x and -x directions, respectively, via selection of appropriate individual magnetic domains.

\begin{figure}
    \centering
    \includegraphics[width = 0.7\columnwidth]{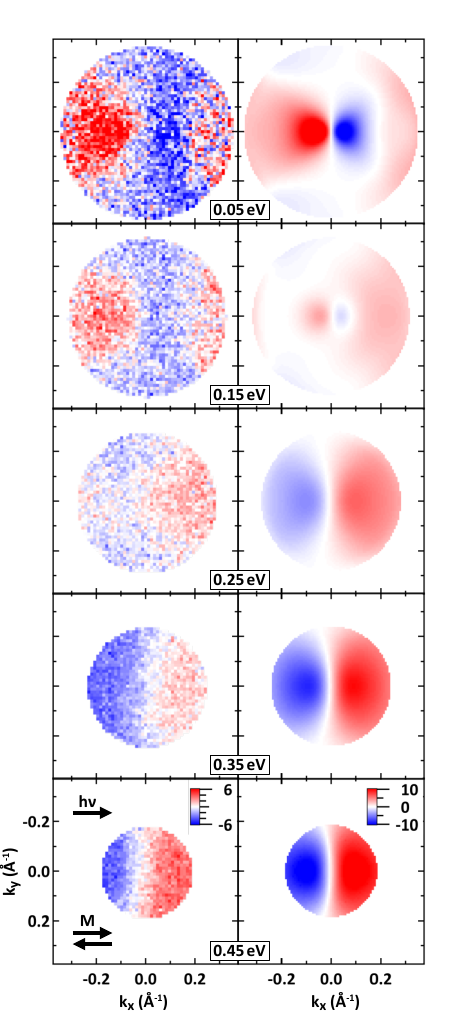}
    \caption{Momentum-resolved exchange asymmetry $A_{\mathrm{ex}}$ of Fe(001) at selected binding energies, as in Fig.~\ref{fig:Apol}. Left column: experimental PEEM data, right column: respective data from  photoemission calculations. Small differences with respect to an odd symmetry upon reversal of $k_{x}$ result from off-normal light incidence. For normal incidence they are absent.}
    \label{fig:Aex}
\end{figure}

The momentum-dependent exchange asymmetry $A_{\mathrm{ex}}$, defined in Eq.~\eqref{eq:Aex} and shown in Fig.~\ref{fig:Aex}, exhibits absolute values up to $\unit[10]{\%}$ in theory and $\unit[6]{\%}$ in experiment, which are an order-of-magnitude stronger effects than previously observed \cite{marx2000}. An odd symmetry of the momentum-dependent $A_{\mathrm{ex}}$ pattern with respect to the $k_{x} = 0$ line would be expected for normal light incidence \cite{schumann2024}. However, for the grazing light incidence here, we find a clear deviation, which results in a curved nodal line between the regions of positive and negative $A_{\mathrm{ex}}$. With respect to the $k_{y} = 0$ line, both experiments and theoretical calculations show a mirror-symmetric pattern in contrast to $A_{\mathrm{pol}}$. The absolute $A_{\mathrm{ex}}$ values, including the sign, depend on the initial state binding energy via the band structure and photoemission matrix elements.

The above findings support that both asymmetries $A_{\mathrm{pol}}$ and $A_{\mathrm{ex}}$ are suitable tools for disentangling and quantifying available contrast mechanisms for darkfield domain imaging. 

From the momentum-resolved $A_{\mathrm{ex}}$ pattern in Fig.~\ref{fig:Aex}, it follows that MCD imaging can selectively reveal strong magnetic contrast in case of \emph{off-normal} electron momentum selection. However, without momentum selection or with a momentum selection centered at $k_x = k_y = 0$, which has been conventionally applied in the literature, different in-plane momentum contributions will largely cancel each other. This cancellation explains the small or vanishing magnetic dichroism for in-plane magnetized domains reported so far.

Our joint experimental and theoretical study suggests to selectively choose the $\vec{k}_{\parallel}$ area of interest in order to enhance the magnetic contrast. Hence, we place a circular contrast aperture in a $\vec{k}_{\parallel}$ area with high exchange asymmetry, a procedure known as darkfield imaging in optics and modern electron microscopy. 

\begin{figure}
    \centering
    \includegraphics[width = 0.7\columnwidth]{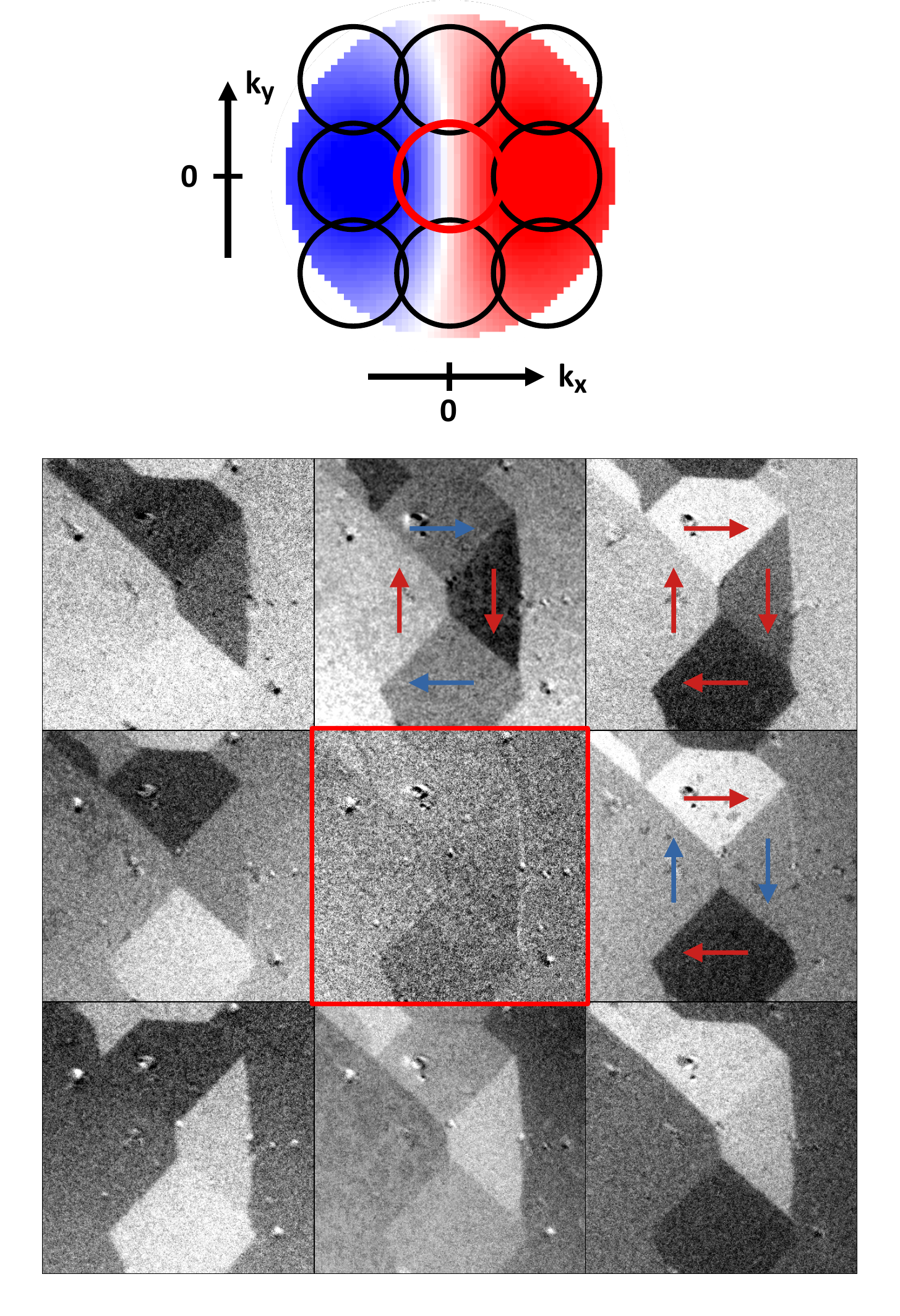}
    \caption{Darkfield MCD imaging of Fe(100). Top: Schematics of the nine aperture positions in the momentum plane, with a momentum-resolved $A_{\mathrm{ex}}$ pattern as background. Bottom: Domain imaging using the nine aperture positions shown above. ($h \nu = \unit[5.2]{eV}$, 
    maximum domain contrast is between 3 and $\unit[4]{\%}$, field of view $\unit[56 \times 56]{\mu m^2}$ each.)}
        %
    \label{fig:Imaging}
\end{figure}

Depending on the position of the aperture the contrast of specific domains is increased, as we show for a Landau-like pattern of four orthogonal magnetic domains at a Fe(001) surface. (Fig.~\ref{fig:Imaging}). Placing the aperture in nine different positions (shown as circles in the top panel of Fig.~\ref{fig:Imaging})  results in nine corresponding MCD PEEM images of the same surface region (bottom panel). 

For the centered aperture, marked in red, the MCD contrast almost vanishes in accordance with our above discussion. However, an aperture centered at $k_x > 0$ and $k_y = 0$ results in a drastically increased contrast of $\unit[3-4]{\%}$ for magnetic domains oriented in $+x$ versus $-x$ direction, whereas the contrast for domains oriented in $+y$ and $-y$ directions (blue arrows) vanishes. Both observations match quantitatively the result of the $\vec{k}_{\parallel}$ space measurements in Fig.~\ref{fig:Aex}. 

Positioning the momentum aperture at $k_x < 0$ and $k_y = 0$ reverses the contrast of $+x$ and $-x$ domains. As expected, the contrast switches from sensitivity in $x$ direction to $y$ direction when positioning the aperture at $k_x = 0$ and $k_y > 0$ (upper-middle PEEM image in Fig.~\ref{fig:Imaging}). The upper-right measurement shows a diagonal position with $k_x > 0$ and $k_y > 0$, where the different contributions to the MCD signal are combined, resulting in four different asymmetry values for the four in-plane magnetization directions. Note that in the latter case also $A_{\mathrm{pol}}$ could contribute besides $A_{\mathrm{ex}}$ to the domain contrasts, however, which is negligible here \cite{schumann2024}. 

\begin{figure}
    \centering
    \includegraphics[width = 0.9\columnwidth]{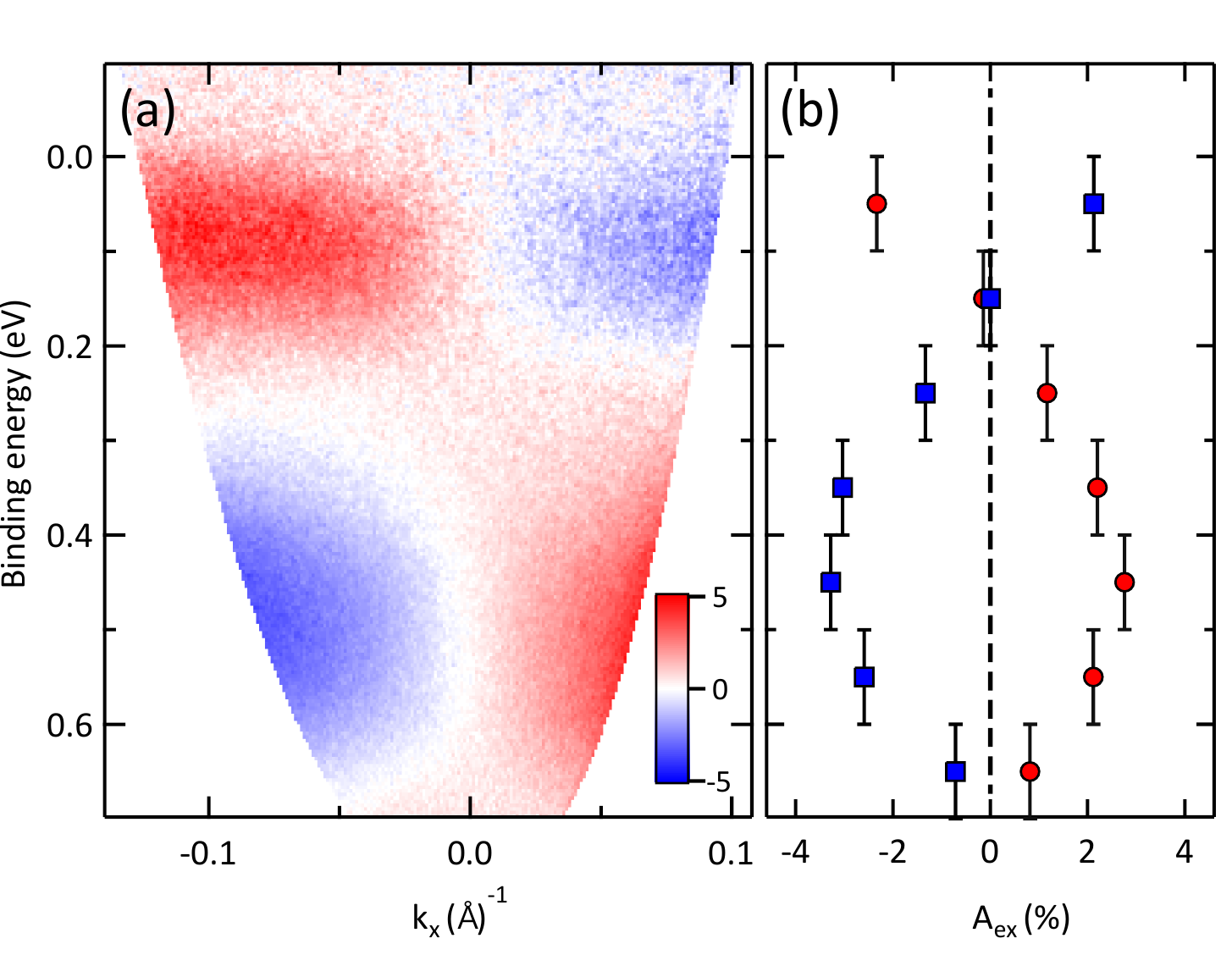}
    \caption{Binding-energy dependent exchange asymmetry $A_{\mathrm{ex}}$ for Fe(001)-(1x1)-O at $h\nu = \unit[5.2]{eV}$. (a) ARPES data for an oxygen-passivated Fe(001) thin film grown on MgO(001) with sample magnetized in $+x$ and $-x$ direction (light incidence at $\unit[70]{^{\circ}}$, $k_y = 0$). (b) Momentum-selected PEEM data for an oxygen-passivated Fe(001) single crystal for positive and negative $k_x$ momentum selection as marked by blue squares and red circles, respectively. (Selection at $|k_x| = \unit[(0.16 \pm 0.12)]{\text{\AA}^{-1}}$, $k_y = \unit[(0 \pm 0.12)]{\text{\AA}^{-1}}$, light incidence at $\unit[65]{^{\circ}}$).}
    \label{fig:AexContrast}
\end{figure}

The magnitude of $A_{\mathrm{ex}}$ and, therefore, of the MCD contrast in PEEM for near-threshold photoemission depend on the initial-state energy, as is demonstrated in Fig.~\ref{fig:Aex}. $A_{\mathrm{ex}}$ reverses sign from up to $\unit[+6]{\%}$ slightly below the Fermi level to $\unit[-4]{\%}$ at $E_{\mathrm{B}} = \unit[0.45]{eV}$. 
As a second, magnetically similar system we studied the oxygen-passivated Fe(001)-(1$\times$1)-O surface with darkfield threshold PEEM, as described above. It yields very similar $A_{\mathrm{pol}}$ and $A_{\mathrm{ex}}$ patterns as those for bare Fe(001) (not shown here), which reverse sign at a binding energy of approximately $\unit[0.2]{eV}$. Momentum-selected $A_{\mathrm{ex}}$ data from momentum-resolved PEEM measurements on 
the two in $\pm x$\, direction magnetized domains are shown in Fig.~\ref{fig:AexContrast}(b) for positive $k_x$ as blue squares and for negative $k_x$ as red circles.

Using an angle-resolved photoelectron spectroscopy (ARPES) setup described previously~\cite{gillmeister2018, gillmeister2020}, the magnetic circular dichroism is analyzed in an independent experiment with higher energy resolution for an $\unit[11]{nm}$ thick Fe(001)-(1$\times$1)-O thin film grown on MgO(001). For fully in $+x$ or $-x$ direction magnetized films, the exchange asymmetry $A_{\mathrm{ex}}$ is depicted in an energy vs momentum map in Fig.~\ref{fig:AexContrast}(a). Note that the acceptance angle of the ARPES spectrometer is limited to $\pm15$\textdegree. Both datasets show large $A_{\mathrm{ex}}$ values, which switch sign upon reversal of $k_x$. At the Fermi level and at $E_{\mathrm{B}} = \unit[0.45]{eV}$ we find strong contrast of about $\unit[5]{\%}$ between $A_{\mathrm{ex}}$ values of +2 and $\unit[-3]{\%}$ with a contrast reversal between 0.15 and $\unit[0.25]{eV}$. We note that this observation requires either precise threshold photoexcitation or energy-resolved electron detection to achieve high MCD signals.
These spectroscopic observations, together with our microscopy data, demonstrate that the full energy-momentum phase space of electronic states, even for paradigmatic systems such as Fe, can be fully utilized for magnetic dichroic domain imaging.

While the darkfield scheme of threshold MCD PEEM is broadly applicable, the magnitude of the binding-energy dependent exchange asymmetry $A_{\mathrm{ex}}$ is a material-specific property. It results from the spin-dependent electronic structure of Fe(001) and the associated ARPES  transition matrix elements. Note that for a fixed binding energy these matrix elements depend on the photon energy due to 
the selective combination of the initial and final electronic states involved.
Note further that our approach can be also implemented into momentum microscopy \cite{Kroemker08}, where the original development stems from back focal plane imaging in a photoelectron microscope \cite{Kotsugi03}.
The ultimate resolution of PEEM-based magnetic imaging depends on the initial resolution of the PEEM instrument in the standard mode as well as an additional reduction for in-plane magnetic imaging due to the off-center electron detection. The latter is unavoidable for darkfield imaging and deteriorates the resolution by a factor between three to nine, depending on the angular photoelectron distribution. 

This study demonstrates that in-plane magnetic domains can be imaged with high contrast using threshold PEEM with momentum selection of the detected photoelectrons, thereby introducing the concept of darkfield threshold MCD PEEM\@. We validated this approach by applying darkfield UV PEEM to an in-plane magnetized Fe(001) surface. However, this method is broadly applicable and can be extended to other ferromagnetic materials, including those with out-of-plane magnetization \cite{kronseder2011}, making it well-suited for investigating magnetic reorientation transitions, such as those observed in Ni/Cu(001) \cite{henk1999,sander2004,nakagawa2006,kronseder2011}.

The most promising potential of this technique lies in its ability to investigate ultrafast magnetization dynamics using femtosecond laser pulses in an optical pump and threshold UV photoemission probe scheme. This capability opens new avenues for studying the ultrafast motion of domain walls \cite{parkin2008} or of large skyrmions on nanometer length scales \cite{goebel2019,jani2021,kern2022}.

\paragraph{Acknowledgments.} 
The authors thank R. Feder for fruitful discussions. We gratefully acknowledge the financial support by the Deutsche Forschungsgemeinschaft (DFG, German Research Foundation) -- Project-ID 328545488 -- TRR~227, projects~A06 and~B04.

\end{document}